\def\,{\thinspace}

\def\kms    {\ifmmode{{\rm \ts km\ts s}^{-1}}\else{\ts km\ts s$^{-1}$}\fi}
\def\kmspc{\kms\,pc$^{-1}$}

\documentclass[oldversion]{aa}
\usepackage{graphicx}
\usepackage{epsfig}
\usepackage{hyperref}

\usepackage{txfonts}
\begin{document}

   \title{Ammonia ($J$,$K$)\,=\,(1,1) to (4,4) and (6,6) inversion lines detected in the Seyfert 2 galaxy NGC 1068}
   \author{Y. Ao \inst{1,2}\thanks{email: ypao@mpifr-bonn.mpg.de}, 
   C. Henkel \inst{1}, J.A. Braatz\inst{3}, A. Wei\ss\inst{1}, K. M. Menten\inst{1} \& S. M\"{u}hle\inst{4}}
   \institute{
   MPIfR, Auf dem H\"{u}gel 69, 53121 Bonn, Germany
   \and
   Purple Mountain Observatory, Chinese Academy of Sciences, Nanjing 210008, China
   \and
   National Radio Astronomy Observatory, 520 Edgemont Rd., Charlottesville, VA 22903, USA
   \and
   Joint Institute for VLBI in Europe, Postbus 2, 7990 AA Dwingeloo, The Netherlands
}
   \date{}

\authorrunning{Y. Ao et al.}
\titlerunning{Ammonia Detected in Seyfert 2 Galaxy NGC 1068}

\abstract{
We present the detection of the ammonia (NH$_3$) ($J$,$K$)\,=\,(1,1) to (4,4)
and (6,6) inversion lines toward the prototypical Seyfert 2 galaxy NGC~1068,
made with the Green Bank Telescope (GBT). This is the first detection of
ammonia in a Seyfert galaxy. The ortho-to-para-NH$_3$ abundance ratio suggests
that the molecule was formed in a warm medium of at least 20\,K.
For the NH$_3$ column density and fractional abundance, we find
(1.09$\pm$0.23)$\times$10$^{14}$~cm$^{-2}$ and (2.9$\pm$0.6)$\times$10$^{-8}$,
respectively, from the inner $\sim$1.2~kpc of NGC~1068.  The kinetic temperature
can be constrained to 80$\pm$20~K for the bulk of the molecular gas, while some
fraction has an even higher temperature of 140$\pm$30~K.

}

\maketitle

\section{Introduction}
\noindent Studies of molecular gas provide information about the gas density,
the temperature, and kinematics within galaxies, help us to understand chemical
evolution over cosmic time and allow us to study the triggering and fueling
mechanisms of star formation and active galactic nuclei (AGN).  Most stars are
formed in dense gas cores, which are embedded in giant molecular clouds (GMCs).
The star-forming activity is related to the dense gas and not to the bulk of
the GMC's material (Gao \& Solomon 2004). Determining the physical properties
of the dense gas in galaxies is therefore of fundamental importance for our
understanding of star formation and the evolution of galaxies.  Among the most
commonly observed species are CO, CS, HCN, and HCO$^+$.  In local dark clouds,
the temperature can be constrained by observations of the $J$\,=\,1$-$0
transition of CO, both because this transition is opaque and thermalized and
because the emission is often extended enough to fill the beam of a single-dish
telescope.  However, in external galaxies, filling factors are much less than
unity. Furthermore, CO, CS, HCN, and HCO$^+$ suffer from a coupled sensitivity
to the kinetic temperature and spatial density, making an observed line ratio
consistent with both a high density at a low temperature and a low density at a
high temperature. Specific information about the individual physical
parameters therefore requires a molecular tracer that possesses a singular
sensitivity to either density or temperature. Ammonia (NH$_3$) is such a
molecule. 

Ammonia is widespread and its level scheme contains inversion
doublets owing to the tunneling of the nitrogen atom through the plane defined
by the three hydrogen atoms. The metastable ($J,K\,=\,J$) rotational levels, for
which the total angular momentum quantum number $J$ is equal to its projection
on the molecule's symmetry axis, are not radiatively but collisionally coupled.
NH$_3$ is therefore a good tracer of the gas kinetic temperature. Another
advantage is that the inversion lines are quite close in
frequency, thus allowing us to measure sources with similar beam sizes and to
use the same telescope-receiver combination, which minimizes calibration
uncertainties of the measured line ratios. As a consequence, NH$_3$ is widely
studied to investigate the physical properties of dark clouds and massive 
star-forming regions in our Galaxy (e.g., Ho \& Townes 1983; Walmsley \& Ungerechts
1983; Bourke et al. 1995; Ceccarelli et al. 2002; Pillai et al. 2006; Wilson et
al. 2006). However, in spite of the high sensitivity and stable baselines of
present state-of-the-art facilities, ammonia multilevel studies in
extragalactic sources are still rare and are limited to the Large Magellanic
Cloud, IC~342, NGC~253, M~51, M~82, Maffei~2, and Arp~220 (Martin \& Ho 1986;
Henkel et al. 2000; Wei\ss\, et al. 2001; Takano et al. 2002; Mauersberger et
al. 2003; Ott et al. 2005, 2010; Takano et al. 2005 ) in the local Universe,
and in absorption to the gravitationally lensed B0218+357 and PKS~1830$-$211 at
redshifts of $z\sim$ 0.7 and 0.9 (Henkel et al. 2005, 2008).

NGC~1068, a prototypical Seyfert 2 galaxy (Antonucci \& Miller 1985), is
located at a distance of 15.5~Mpc (a heliocentric systemic velocity
$cz$\,=\,1137~\kms\, is used throughout this paper, adopted from the NASA/IPA
Extragalactic Database), making it one of the nearest Seyfert galaxies and thus
an ideal target to investigate the physical properties of the molecular gas in
the vicinity of an AGN.  Here we report the first detection of ammonia in this
prominent Seyfert 2 galaxy to evaluate the kinetic temperature of its dense
molecular gas.

\section{Observations}\label{observation}
We observed ammonia inversion lines toward the nucleus of NGC 1068 with the
Green Bank Telescope (GBT) of the National Radio Astronomy
Observatory\footnote{The National Radio Astronomy Observatory is a facility of
the National Science Foundation operated under cooperative agreement with
Associated Universities, Inc.} on 2005 October 14 and October 19.  We
configured the telescope to observe both circular polarizations simultaneously
using a total-power nodding mode in which the target position was placed
alternately in one of the two beams of the 22$-$26 GHz K-band receiver. The
full-width-at-half-maximum (FWHM) beam size was approximately 31$\arcsec$ in the
observed frequency range, between 23.6 and 25.0 GHz.  The pointing accuracy was
$\sim$~5$\arcsec$. To observe the NH$_3$(1,1) through (4,4) lines, we
configured the spectrometer with an 800 MHz spectral window centered on the
NH$_3$(3,3) line. The channel spacing was 390 kHz, corresponding to
$\sim$~5~\kms. We averaged the data from the two dates to produce the final
spectrum, which has an on-source integration time of 4 hours.  We observed the
NH$_3$(6,6) line only on the second date, 2005 October 19.  Here we used a 200
MHz spectral window observed with a channel spacing of 24 kHz, and then
averaged channels in post-processing to achieve 390 kHz channel spacing.  The
total integration time for the NH$_3$(6,6) line was 1 hour.

The data were calibrated and averaged in GBTIDL. For calibration, we used an
estimate of the atmospheric opacity obtained from a weather model.  We
subtracted a polynomial baseline fitted to the line-free channels adjacent to
each ammonia line. For the (1,1) to (4,4) lines observed simultaneously,
the relative calibration is very good and is limited primarily by the uncertainty
in the baseline fits, which is 10\% for the (1,1) to (3,3) lines, 15\% for the
(4,4) line, and 27\% for the (6,6) line, respectively. The (1,1) and (2,2) line
profiles are close to each other but are marginally separated and the Gaussian
fit for each line does not suffer from this effect. With the calibration error
of 15\% for the GBT telescope itself, we estimate the absolute calibration
uncertainties, including the error from the baseline fits, to be $\sim$ 18-31\%
for the ammonia lines.

\section{Results}
\subsection{NH$_3$ lines}\label{ammonia}
\begin{figure}[t]
\vspace{-0.0cm}
\centering
\includegraphics[angle=0,width=0.5\textwidth]{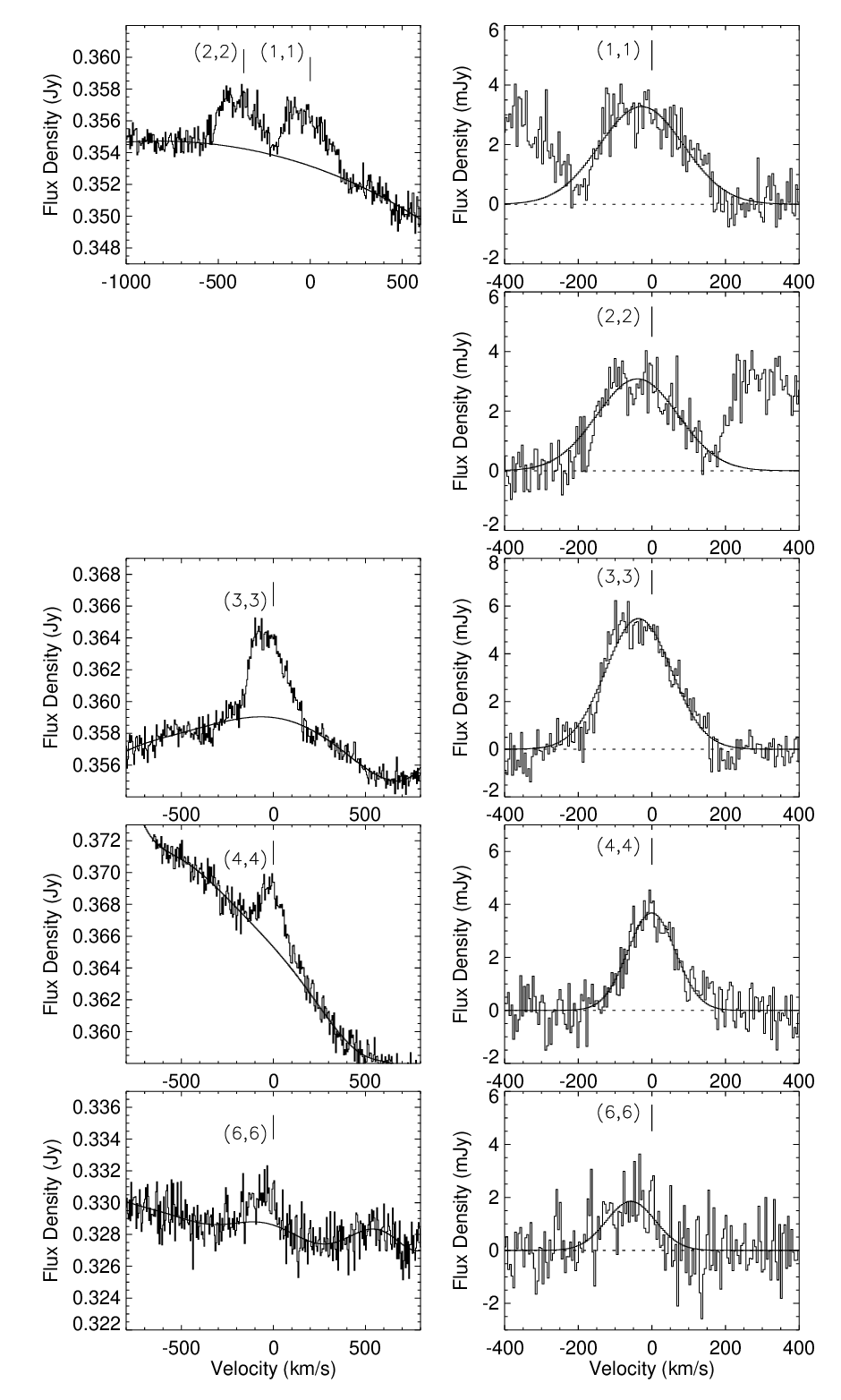}
\vspace{-0.0cm}
\caption{Metastable ammonia inversion lines observed toward NGC~1068
($\alpha_{\rm 2000}$\,=\,02$\rm^h$42$\rm^m$40.7$\rm^s$, $\delta_{\rm
2000}$\,=\,$-$00$\rm^o$00$\arcmin$48$\arcsec$).  Vertical lines mark a
heliocentric systemic velocity of $cz$\,=\,1137~\kms. Velocities are displayed
relative to this value (in the upper left panel, the velocity scale refers to
the (1,1) line). {\it Left:} Spectra without baseline subtraction, overlaid
with a polynomial baseline fit to the line-free channels.  {\it Right:} Final
baseline-subtracted spectra overlaid with Gaussian fits.}\label{spectra}
\end{figure}

This is the first time that ammonia is detected in NGC~1068, making it
the first Seyfert galaxy observed in this molecule.  Also, because Arp~220,
B0218+357, and PKS~1830-211 were detected in absorption, NGC~1068 is so far the
most distant galaxy where ammonia is detected in emission. Three para,
($J,K$)\,=\,(1,1), (2,2), and (4,4), and two ortho, ($J,K$)\,=\,(3,3) and
(6,6), transitions are covered by the observations, and all five are
detected.  Figure~\ref{spectra} shows the spectra before and after
subtracting a polynomial baseline to the line free channels. The line
parameters are listed in Table~\ref{tab1}, where integrated flux, 
$I$, peak flux density, $S$, central velocity, $V$, and line width, 
$\Delta V_{\rm 1/2}$, are obtained from the results of Gaussian fitting.

The line profiles are similar for the (1,1), (2,2) and (3,3) lines, which share
a similar central velocity of around $-$35~\kms, relative to a heliocentric
systemic velocity of $cz$\,=\,1137~\kms, and a FWHM line width of
220$-$274~\kms. The similar central velocities and the comparable widths of the
(1,1), (2,2) and (3,3) lines suggest that the line emission comes from the same
region. The (4,4) and (6,6) lines have different central velocities and a
narrower line width.  The (4,4) line is located at a frequency where the
baseline is particularly steep.  The (6,6) line is weak and falls into a window
with a slightly distorted baseline, making it difficult to determine accurate
line parameters. The comparatively narrow line widths of the (4,4) and (6,6)
lines may suggest that their emission originates from a less extended region
than that of the lower excitation lines.

Our measured continuum flux density of NGC~1068 ranges from 0.33 to 0.37 Jy at
frequencies of 23.4 to 25.1 GHz. This is consistent with
the value of 0.342$\pm$0.034 at 22 GHz reported by Ricci et al. (2006).

\begin{center}
\begin{table*}
\caption[]{NH$_3$ line parameters}\label{tab1}
\begin{flushleft}
\begin{tabular}{llllll}
\hline
(J,K)  & $I$$^a$  & $S$$^a$  & $V^a$  &  $\Delta V_{\rm 1/2}$$^a$   & $N(J,K)$$^b$  \\
            & (Jy\,km\,s$^{-1}$)   & (mJy)  & (\kms)  &   (\kms)   &  (10$^{13}$ cm$^{-2}$) \\
\hline
(1,1) &  0.89( 0.07) &  3.26( 0.13) &  $-$30(6)  & 274(17) &  2.52(0.24) \\
(2,2) &  0.79( 0.06) &  3.10( 0.14) &  $-$37(6)  & 254(19) &  1.67(0.19) \\
(3,3) &  1.19( 0.10) &  5.40( 0.14) &  $-$37(3)  & 220( 8) &  2.22(0.13) \\
(4,4) &  0.55( 0.05) &  3.72( 0.14) &   $-$2(3)  & 149( 7) &  0.96(0.07) \\
(6,6) &  0.28( 0.04) &  1.88( 0.23) &  $-$56(9)  & 147(27) &  0.43(0.12) \\
(5,5) &              &              &  &         &  0.53 \\
(0,0) &              &              &  &         &  2.55 \\
$N_{\rm total}$ &              &              &  &         &  10.9(2.3) \\
\hline
\end{tabular}
\begin{list}{}{}
\item{$^a$ The values are derived from Gaussian fits to the spectra.  Given
velocities are relative to a heliocentric velocity of $cz$\,=\,1137~\kms.}
\item{$^b$ The column density for the (5,5) state is extrapolated from the
(2,2) and (4,4) states using $T_{\rm rot}$\,=\,119~K, while the value for the
(0,0) state is extrapolated from the (3,3) state by adopting $T_{\rm
rot}$\,=\,44~K. The total column density, $N_{\rm total}$, includes the
populations of the levels from (0,0) to (6,6).  Because relative calibration is
excellent up to at least the (4,4) transition, the errors given in parenthesis
refer exclusively to the Gaussian fits.  For the cumulative column density,
$N_{\rm total}$, however, the given error accounts for the absolute
calibration uncertainty ($\S$~\ref{observation}).}
\end{list}
\end{flushleft}
\end{table*}
\end{center}
\subsection{NH$_3$ column density and rotation temperature}\label{trot}
\begin{figure}[t]
\vspace{-0.0cm}
\centering
\includegraphics[angle=0,width=0.45\textwidth]{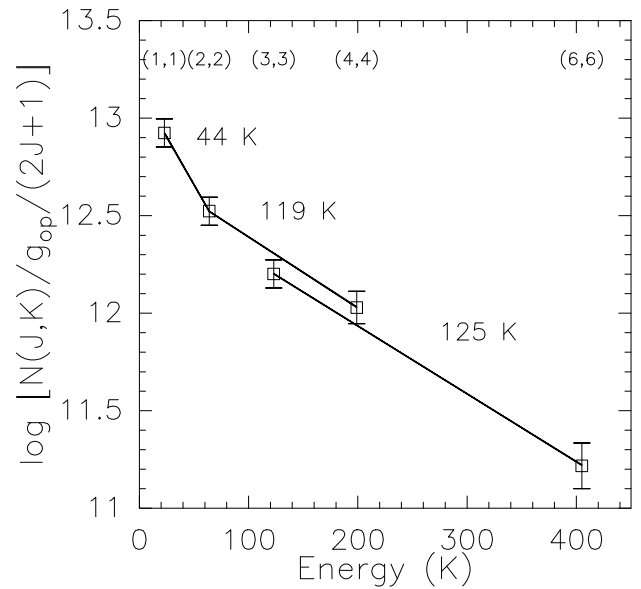}
\vspace{-0.0cm}
\caption{Rotation diagram of metastable ammonia transitions toward NGC~1068
(see \S~\ref{trot}). The open squares show the normalized column densities
determined from the integrated line intensities.  The numbers mark the
rotational temperatures in K.  The absolute calibration uncertainties,
including the dominant contribution from the baseline fits as well as
uncertainties in the Gaussian fits and in the overall calibration uncertainty,
have been taken as error bars ($\S$~\ref{observation}).  Note, however, that
relative calibration is excellent up to at least the (4,4) transition.
}\label{bmplot}
\end{figure}

Assuming that the line emission is optically thin and the contribution from the
cosmic background is negligible, the sum of the beam-averaged column densities
of the two states of an inversion doublet can be calculated using
\begin{equation} N(J,K)\,=\,\frac{\rm 1.55 \times
10^{14}}{\nu}\frac{J(J+1)}{K^2}\int T_{\rm mb}{\rm d v} \end{equation} (e.g.,
Mauersberger et al. 2003), where the column density $N(J,K)$, the frequency
$\nu$, and the integrated line intensity $\int T_{\rm mb}{\rm d v}$, based on the main
beam brightness temperature, $T_{\rm mb}$, are in units of cm$^{-2}$, GHz, and
K\,\kms, respectively.  The calculated column densities (1~Jy corresponds to
2.16~K on a $T_{\rm mb}$ scale\footnote{for the details see the memo of
``Calibration of GBT Spectral Line Data in GBTIDL''
\href{http://www.gb.nrao.edu/GBT/DA/gbtidl/gbtidl_calibration.pdf}{$\rm
{http://www.gb.nrao.edu/GBT/DA/gbtidl/gbtidl\_calibration.pdf}$}}) are given in
Table~\ref{tab1}.

Following the analysis described by Henkel et al. (2000), the rotational
temperature ($T_{\rm rot}$) between different energy levels can be determined
from the slope, $a$, of a linear fit in the rotation diagram (i.e., Boltzmann
plot, which is normalized column density against energy above the ground state
expressed in $E/k$) by $T_{\rm rot}$\,=\,$-{\rm
log}\,e/a$$\,\approx\,-0.434/a$. Figure~\ref{bmplot} shows the rotation diagram
including the five measured metastable NH$_{\rm 3}$ inversion lines. 
Ignoring differences in line shape, we obtain $T_{\rm
rot}$\,=\,92$^{+29}_{-18}$~K by fitting the para-NH$_3$ species, i.e.\ the
(1,1), (2,2) and (4,4) transitions, and $T_{\rm rot}$\,=\,125$^{+11}_{-8}$~K
for the (3,3) and (6,6) ortho-NH$_3$ transitions.
As discussed in $\S~\ref{ammonia}$, the (1,1), (2,2) and (3,3) line emission may
originate in a region that is different from that of the (4,4) and (6,6)
lines. The former three lines can be fitted by $T_{\rm
rot}$\,=\,61$^{+12}_{-9}$~K, and the latter two by $T_{\rm
rot}$\,=\,111$^{+12}_{-9}$~K. The rotation temperature between the lowest
inversion doublets of para-ammonia, (1,1) and (2,2), is $T_{\rm
rot}$\,=\,44$^{+6}_{-4}$~K. Without the (1,1) state, the other two para-NH$_3$
transitions, (2,2) and (4,4), yield $T_{\rm rot}$\,=\,119$^{+15}_{-11}$~K.

\subsection{NH$_3$ abundance}\label{abundance}
To estimate the populations of the $(J,K)$\,=\,(0,0) and (5,5) states, we
assume the rotation temperature $T_{\rm rot}$\,=\,44~K, derived from the two
lowest inversion doublets, for the (0,0) ground state, and $T_{\rm
rot}$\,=\,119~K from the fit to the (2,2) and (4,4) lines, for the (5,5) state.
The derived column densities are 2.55$\times$10$^{13}$ cm$^{-2}$ for the (0,0)
state, which is not a doublet, and 0.53$\times$10$^{13}$ cm$^{-2}$ for the
(5,5) state, respectively. The former is extrapolated from the (3,3) state and
the latter from the (2,2) and (4,4) states. The resulting total column density
of ammonia is (1.09$\pm$0.23)$\times$10$^{14}$ cm$^{-2}$ if only considering
the populations in the metastable states up to ($J,K$)\,=\,(6,6).  This yields
an ammonia gas mass of 68$\pm$14~M$_\odot$ within a beam size of 31$\arcsec$,
which corresponds to a physical size of 2.3~kpc.  To estimate the mass of
molecular gas in NGC~1068, we convolved the 17$\arcsec$ resolution map in CO
$J$\,=\,1$-$0 by Kaneko et al. (1989) to the GBT's resolution of 31$\arcsec$
and derived an integrated intensity of 103~K~\kms.  Adopting a conversion
factor of 0.8~M$_\odot$~${(\rm K\,km\,s^{-1}\,pc^2)^{-1}}$, which describes
ultraluminous galaxies (Downes \& Solomon 1998) and less conspicuous nuclear
starbursts (Mauersberger et al. 2003), we obtain a molecular gas mass of
3.5$\times$10$^{8}$~M$_\odot$. Therefore, the ammonia abundance relative to
molecular hydrogen is estimated to be (2.9$\pm$0.6)$\times$10$^{-8}$ within a
radius of $\sim$1.2~kpc around the center of NGC~1068.

\section{Discussion}\label{discussion}
\begin{figure}[t]
\vspace{-0.0cm}
\centering
\includegraphics[angle=0,width=0.45\textwidth]{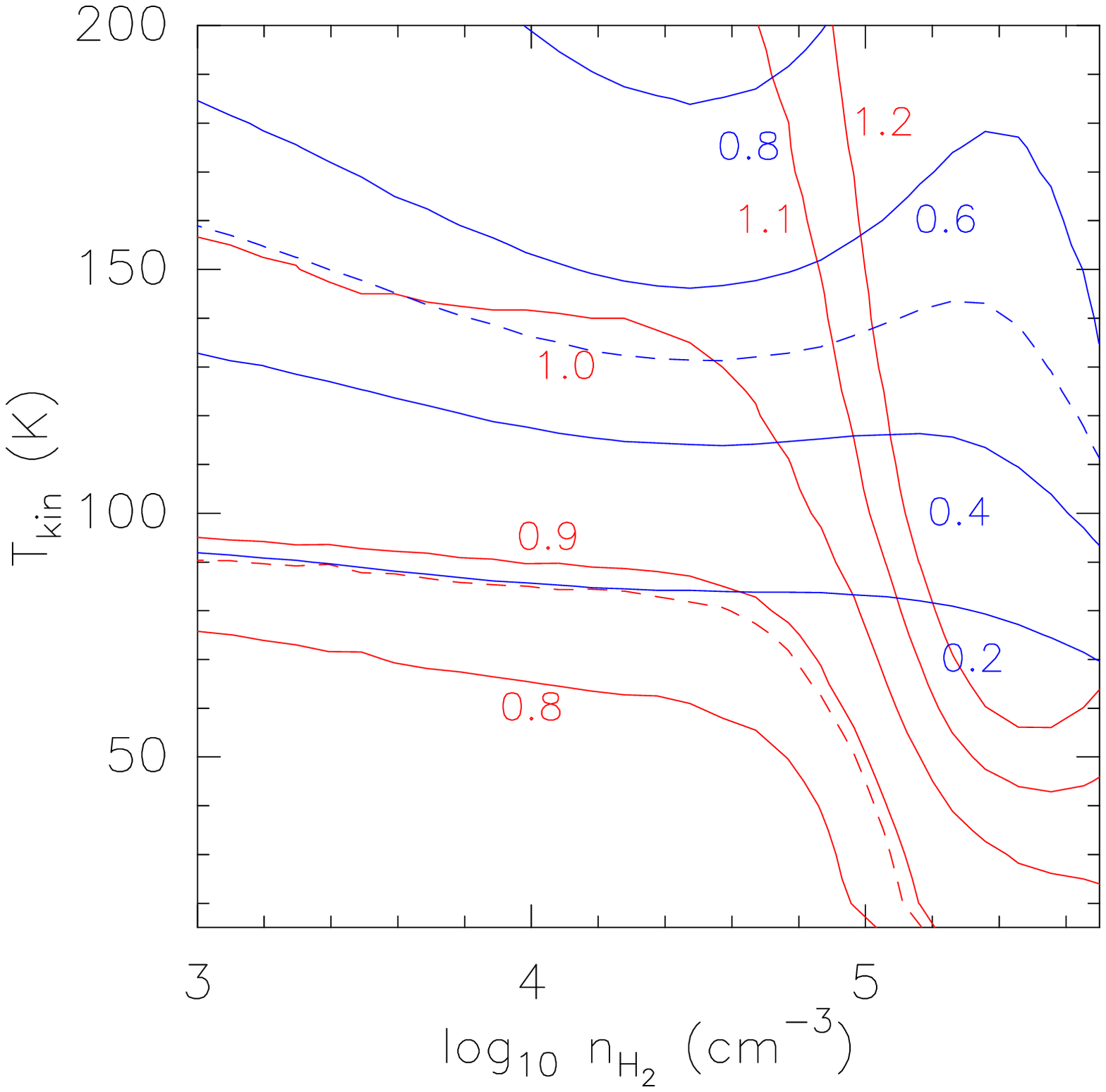}
\vspace{-0.0cm}
\caption{
NH$_3$ inversion line ratios as a function of the H$_2$ density ($n_{\rm H_2}$)
and kinetic temperature ($T_{\rm kin}$). The red and blue lines mark the
(2,2)/(1,1) and (6,6)/(4,4) line ratios, respectively, obtained from the large
velocity gradient model outlined in \S~\ref{discussion}. Dashed lines show the
line ratios measured with the GBT. [NH$_3$]/(d$v$/d$r$)\,=\,10$^{-8}$ pc~(km
s$^{-1}$)$^{-1}$ ([NH$_3$]\,=\,N(NH$_3$)/N(H$_2$) is the fractional abundance
of ammonia).  At high densities the (1,1) and (2,2) lines saturate, which
causes the steeply declining $T_{\rm kin}$ values with rising density for a
given (2,2)/(1,1) line ratio. }\label{tkin}
\end{figure}

The ammonia abundance in NGC~1068 is consistent with the values of
(1.3$-$2.0)$\times$10$^{-8}$ reported by Mauersberger et al. (2003) for the
central regions of several nearby galaxies such as NGC~253, Maffei~2 and
IC~342.  It is lower than that of a few $\times$10$^{-7}$ for the central
$\sim$ 500~pc of the Milky Way (Rodr{\'{\i}}guez-Fern{\'a}ndez et al. 2001),
but higher than the extremely low value of 5$\times$10$^{-10}$ determined for
M~82 by Wei\ss\, et al. (2001).  M~82 is a special case, because it contains a starburst
in a late stage of its evolution leaving NH$_3$, which is a molecule
particularly sensitive to UV radiation (e.g., Suto \& Lee 1983), in only a few
molecular cores that are still sufficiently shielded against the intense radiation
field. 

With the three para- and two ortho-NH$_3$ lines observed, we can estimate the
ortho-to-para abundance ratio to be $R$ $\sim$ 0.9 (we do not give an error
here because of the extrapolated (0,0) column density) as an upper limit
because the (0,0) column density might be overestimated by extrapolating from
the (3,3) column density, which could be affected by maser activity (e.g.,
Walmsley \& Ungerechts 1983; Ott et al. 2005). A value near or below
unity is expected, if ammonia was formed in a warm medium at a gas kinetic
temperature of at least 20\,K and only $R$ values above $\sim$1.5 would hint at
a cool formation temperature of $\sim$ 20~K (e.g., Takano et al. 2002).

The rotation temperatures from the multilevel study of the ammonia inversion
lines are a good approximation to the kinetic gas temperature only for low
($T_{\rm kin}$ $\le$ 20~K) temperatures (Walmsley \& Ungerechts 1983; Danby et
al. 1988). At higher temperatures, the rotation temperature provides a robust
lower limit to the kinetic gas temperature as long as saturation effects do not
play a role.

To determine the kinetic temperature itself, we here use a one-component large
velocity gradient (LVG) analysis adopting a spherical cloud geometry as
described by Ott et al. (2005).  Dahmen et al. (1998) estimated that the
velocity gradient ranges from 3 to 6~\kmspc\, for Galactic center clouds and
Meier et al. (2008) found a typical value of 1 to 2~\kmspc\, for GMCs. Here we
adopt a median value of 3~\kmspc\, for the LVG models presented in this paper.
The ammonia abundance derived in $\S$~\ref{abundance} is adopted, which yields
an NH$_3$ abundance per velocity gradient of [NH$_3$]/(d$v$/d$r$) $\sim$
10$^{-8}$ pc~(km s$^{-1}$)$^{-1}$ ([NH$_3$]\,=\,N(NH$_3$)/N(H$_2$) is the
fractional abundance of ammonia). Changing the velocity gradient by a factor
of 3 will affect the derived kinetic temperature by less than 10~K.
Figure~\ref{tkin} shows NH$_3$ inversion line ratios as a function of the H$_2$
density ($n_{\rm H_2}$) and kinetic temperature ($T_{\rm kin}$). As long as the
NH$_3$ lines are optically thin, the ratios are almost independent of the gas
density and are therefore a good indicator for gas kinetic temperature.  At high
densities the (1,1) and (2,2) lines saturate, which causes the steeply
declining $T_{\rm kin}$ values with rising density for a given (2,2)/(1,1) line
ratio.  While the optical depth may be well above unity in molecular cores
(see, G{\"u}sten et al. 1981 for the Galactic center region), the bulk of the
ammonia emission should be optically thin and arises from gas densities $n_{\rm
H_2}$\,$<$\,10$^5$ cm$^{-3}$ (e.g., Tieftrunk et al. 1998). For a typical gas
density, $n_{\rm H_2}$, of 10$^{3.0-4.8}$~cm$^{-3}$, the observed (2,2)/(1,1)
line ratio constrains the kinetic temperature $T_{\rm kin}$ to 80$\pm$20~K,
which characterizes the bulk of the molecular gas. The (6,6)/(4,4) line ratio
yields a gas temperature of 140$\pm$30~K for a gas density within the range of
10$^{3.0-5.5}$~cm$^{-3}$. The latter indicates the existence of a hotter
component of the gas, as is also suggested by the high rotational temperature
revealed by the rotation diagram (Fig.~\ref{bmplot}). Here the (6,6)/(3,3)
line ratio is not used to estimate the temperature because the
(3,3) line emission is partly affected by the maser activity in some parameter ranges
(Walmsley \& Ungerechts 1983), and because the (6,6) line was not observed
simultaneously as the (3,3) line and the uncertainty of the baseline fit
of the (6,6) line is largest among all spectra as described in $\S~\ref{observation}$.

The GBT beam size covers the two inner spiral arms and the circumnuclear disk
(CND) of NGC~1068 (see, e.g., Schinnerer et al. 2000). Our observed ammonia
line emission likely has contributions from the spiral arms and the CND.
The CND may be warmer because it is heated not only by young stars but also by
the AGN.  There are observations that directly support higher temperatures in
the CND. Infrared rotational transitions of molecular hydrogen (Lutz et al.
1997) indicate a wide range of gas temperatures between 100 and 800~K. The dust
near the AGN has been particularly thoroughly studied (e.g., Tomono et al.
2006; Poncelet et al. 2007; Raban et al. 2009). Because densities may be high
within the central arcsecond ($\ga$\,10$^5$ cm$^{-3}$), the dust and the gas
phase may be coupled and thus these results may be relevant for the kinetic
temperature of the gas component.  With mid-infrared (MIR) multi-filter data,
Tomono et al. (2006) obtained a dust temperature of $\sim$200~K within a
1.0$\arcsec$-sized region ($\sim$80~pc).

Our data provide a direct estimate of the gas properties for molecular gas in
the inner 1.2~kpc of the galaxy.  However, higher angular resolution data are
needed to separate the different components and to isolate the CND in this
galaxy.

\section{Conclusions}
Our main results from the ammonia observations toward the prototypical Seyfert 2 galaxy
NGC~1068 are:

(1) The metastable NH$_3$ ($J$,$K$)\,=\,(1,1) to (4,4) and (6,6) inversion
lines are for the first time detected in emission toward NGC~1068. 
This opens up a new avenue to determine kinetic temperatures of the
dense gas in nearby highly obscured AGN.

(2) For the NH$_3$ column density and fractional abundance, we find
(1.09$\pm$0.23)$\times$10$^{14}$~cm$^{-2}$ and (2.9$\pm$0.6)$\times$10$^{-8}$
in the inner $\sim$1.2~kpc of NGC~1068.

(3) With an ortho-to-para-NH$_3$ abundance ratio of $\sim$0.9, the ammonia should
have been formed in a warm medium of at least 20\,K.
 
(4) The kinetic temperature can be constrained to 80$\pm$20~K for the bulk of
the molecular gas, while some gas fraction has an even higher temperature of
140$\pm$30~K.


\begin{acknowledgements}
We thank the referee for thoughtful comments that improved this paper and the
staff at the GBT for their supporting during the observations. Y.A.
acknowledges the supports by CAS/SAFEA International Partnership Program for
Creative Research Teams (No. KTCX2-YW-T14), grant 11003044 from the National
Natural Science Foundation of China, and 2009's President Excellent Thesis
Award of the Chinese Academy of Sciences. This research has made use of NASA’s
Astrophysical Data System (ADS).

\end{acknowledgements}


\begin{thebibliography}{}

\bibitem[1985]{1985} Antonucci, R.~R.~J., \& Miller, J.~S.\ 1985, \apj, 297, 621 

\bibitem[1995]{1995} Bourke, T.~L., Hyland, A.~R., Robinson, G., James, 
S.~D., \& Wright, C.~M.\ 1995, \mnras, 276, 1067 


\bibitem[2002]{2002} Ceccarelli, C., Baluteau, J.-P., Walmsley, M., et al.\ 
2002, \aap, 383, 603 


\bibitem[1998]{1998} Dahmen, G., Huttemeister, S., Wilson, T.~L., 
\& Mauersberger, R.\ 1998, \aap, 331, 959 


\bibitem[1988]{1988} Danby, G., Flower, D.~R., Valiron, P., Schilke, P., 
\& Walmsley, C.~M.\ 1988, \mnras, 235, 229 


\bibitem[1998]{1998} Downes, D., \& Solomon, P.~M.\ 1998, \apj, 507, 615 


\bibitem[2004]{2004} Gao, Y., \& Solomon, P.~M.\ 2004, \apj, 606, 271 


\bibitem[1981]{1981} Guesten, R., Walmsley, C.~M., 
\& Pauls, T.\ 1981, \aap, 103, 197 


\bibitem[2008]{2008} Henkel, C., Braatz, J.~A., Menten, K.~M., 
\& Ott, J.\ 2008, \aap, 485, 451 


\bibitem[2005]{2005} Henkel, C., Jethava, N., Kraus, A., et al.\ 2005, 
\aap, 440, 893 


\bibitem[2000]{2000} Henkel, C., Mauersberger, R., Peck, A.~B., Falcke, H., 
\& Hagiwara, Y.\ 2000, \aap, 361, L45 


\bibitem[1983]{1983} Ho, P.~T.~P., \& Townes, C.~H.\ 1983, \araa, 21, 239 


\bibitem[1989]{1989} Kaneko, N., Morita, K., Fukui, Y., et al.\ 1989, \apj, 
337, 691 


\bibitem[1997]{1997} Lutz, D., Sturm, E., Genzel, R., Moorwood, A.~F.~M., 
\& Sternberg, A.\ 1997, \apss, 248, 217 


\bibitem[1986]{1986} Martin, R.~N., \& Ho, P.~T.~P.\ 1986, \apjl, 308, L7 


\bibitem[2003]{2003} Mauersberger, R., Henkel, C., Wei{\ss}, A., Peck, 
A.~B., \& Hagiwara, Y.\ 2003, \aap, 403, 561 


\bibitem[2008]{2008} Meier, D.~S., Turner, J.~L., 
\& Hurt, R.~L.\ 2008, \apj, 675, 281 


\bibitem[2010]{2010} Ott, J., Henkel, C., Staveley-Smith, L., 
\& Wei{\ss}, A.\ 2010, \apj, 710, 105 


\bibitem[2005]{2005} Ott, J., Weiss, A., Henkel, C., 
\& Walter, F.\ 2005, \apj, 629, 767 


\bibitem[2006]{2006} Pillai, T., Wyrowski, F., Carey, S.~J., 
\& Menten, K.~M.\ 2006, \aap, 450, 569 


\bibitem[2007]{2007} Poncelet, A., Doucet, C., Perrin, G., Sol, H., 
\& Lagage, P.~O.\ 2007, \aap, 472, 823 


\bibitem[2009]{2009} Raban, D., Jaffe, W., R{\"o}ttgering, H., 
Meisenheimer, K., \& Tristram, K.~R.~W.\ 2009, \mnras, 394, 1325 


\bibitem[2006]{2006} Ricci, R., Prandoni, I., Gruppioni, C., Sault, R.~J., 
\& de Zotti, G.\ 2006, \aap, 445, 465 


\bibitem[2000]{2000} Schinnerer, E., Eckart, A., Tacconi, L.~J., Genzel, 
R., \& Downes, D.\ 2000, \apj, 533, 850 

\bibitem[Suto \& 1983]{1983} Suto, M., \& Lee, L. C. 1983, J. Chem. Phys., 78, 4515

\bibitem[2005]{2005} Takano, S., Hofner, P., Winnewisser, G., Nakai, N., 
\& Kawaguchi, K.\ 2005, \pasj, 57, 549 


\bibitem[2002]{2002} Takano, S., Nakai, N., 
\& Kawaguchi, K.\ 2002, \pasj, 54, 195 


\bibitem[1998]{1998} Tieftrunk, A.~R., Megeath, S.~T., Wilson, T.~L., 
\& Rayner, J.~T.\ 1998, \aap, 336, 991 


\bibitem[2006]{2006} Tomono, D., Terada, H., 
\& Kobayashi, N.\ 2006, \apj, 646, 774 


\bibitem[1983]{1983} Walmsley, C.~M., 
\& Ungerechts, H.\ 1983, \aap, 122, 164 


\bibitem[2001]{2001} Wei{\ss}, A., Neininger, N., Henkel, C., Stutzki, J., 
\& Klein, U.\ 2001, \apjl, 554, L143 


\bibitem[2006]{2006} Wilson, T.~L., Henkel, C., \& H{\"u}ttemeister, S.\ 2006, \aap, 460, 533 


\end{thebibliography}
\end{document}